\documentclass[epj]{svjour}

\usepackage{graphics}
\usepackage{siunitx}
\usepackage{amsmath}
\usepackage{amsmath,amssymb}
\usepackage{times}
\usepackage{textcomp}
\usepackage{xcolor}
\usepackage{color}
\begin{document}
\titlerunning{Probing roto-translational diffusion with a bright-field microscope}
\title{Probing roto-translational diffusion of small anisotropic colloidal particles with a bright-field microscope}

\author{Fabio Giavazzi\inst{1}\thanks{Corresponding author: fabio.giavazzi@unimi.it}, Antara Pal\inst{2},\and Roberto Cerbino\inst{1,3}
}                     
%
%
\institute{Dipartimento di Biotecnologie Mediche e Medicina Traslazionale, Universit\`a degli Studi di Milano, via F.lli Cervi 93, 20090 Segrate, Italy \and Division of Physical Chemistry, Department of Chemistry, Lund University, Lund, Sweden \and University of Vienna, Faculty of Physics, Boltzmanngasse 5, 1090 Vienna, Austria}

\date{Received: date / Revised version: date}
%

\abstract{
Soft and biological materials are often composed of elementary constituents exhibiting an incessant roto-translational motion at the microscopic scale. Tracking this motion with a bright-field microscope becomes increasingly challenging when the particle size becomes smaller than the microscope resolution, a case which is frequently encountered. Here we demonstrate Squared-Gradient Differential Dynamic Microscopy (SG-DDM) as a tool to successfully use bright-field microscopy to extract the roto-translational dynamics of small anisotropic colloidal particles, whose rotational motion cannot be tracked accurately in direct space. We provide analytical justification and experimental demonstration of the method by successful application to an aqueous suspension of peanut-shaped particles.}
\maketitle
\section{Introduction}
Over a century ago, Jean Perrin conducted a series of experiments to mark the birth of quantitative microscopy \cite{perrin2013brownian,cerbino2018quantitative}. Among his many intriguing results, the simultaneous measurement of the translational and rotational diffusion coefficients of micron-sized particles provided the first convincing experimental demonstration of the energy equipartition between the translational and rotational degrees of freedom, one of the most important predictions of the kinetic theory.
Since then, the rotational and the roto-translational motion of colloidal objects has been repeatedly used as the tool of choice to investigate fundamental physical processes like mass transport or phase transitions \cite{han2006brownian,edmond2012decoupling}, as a probe of material properties at the micro-scale
\cite{PhysRevLett.90.018304,colin2014rotational},
as a biosensor \cite{chen2020trace},
to monitor the motility pattern of active swimmers and their interactions with passive particles \cite{saragosti2012modeling,peng2016diffusion}, to characterize complex transport dynamics in cells \cite{gao2017seeing}, just to mention a few recent examples.

Nowadays, optical microscopy remains a key tool for probing roto-translational particle motion, especially in combination with specific (for example fluorescent) markers. For low particle density, single particle tracking (SPT) represents a rather straightforward way to extract a quantitative information on the translational dynamics of well-resolved particles by following their center of mass; by contrast, the accurate determination of their three-dimensional rotational dynamics with SPT is usually much more challenging and time consuming \cite{anthony2015tracking}. Moreover, most of the SPT algorithms are developed and optimized for spherical particles. 

The study of the roto-translational dynamic of small anisotropic particles and/or of particles at high densities is thus the realm of ensemble-averaging techniques, such as depolarized dynamic light scattering (DDLS)\cite{berne2000dynamic,PhysRevE.52.2707}, polarized fluorescence recovery after photobleaching (pFRAP) \cite{velez1988polarized,lettinga2004rotational}, and nuclear magnetic resonance (NMR). However, the main limitation or these approaches is that they cannot be used in complex, highly heterogeneous environments, which is a common scenario in biology and material science, and one for which microscopy is particularly utilized.

An intermediate approach is provided by a family of methods known as Digital Fourier microscopy (DFM) \cite{giavazzi2014digital}: as in microscopy, temporal image sequences of the sample are acquired in direct space but the sample intermediate scattering function $f(q,\tau)$ is extracted via image correlation in the reciprocal (Fourier) space, as a function of the scattering wave-vector $q$ and of the time delay $\tau$. One of the most known implementation of DFM is Differential Dynamic Microscopy (DDM), a technique that has found vast application with a variety of soft and biological matter systems \cite{cerbino2008differential,cerbino2017perspective}. DDM has been successfully used for studying the translational dynamics of shape- or functionally-anisotropic particles \cite{reufer2012differential,PhysRevLett.121.078001,nixon2019differential,pal2020anisotropic}. In combination with a suitable imaging mode, like polarized \cite{giavazzi2016simultaneous,edera2017differential} or dark-field \cite{cerbino2017dark} microscopy, DDM enables the accurate measurement of the roto-translational dynamics of optically and/or shape-anisotropic particles.

While measuring the translational motion in DDM is based on the simple fact that a translating particle occupies different portions of an image, the key for DDM to be able to capture the particle rotational motion is to convert the latter into an intensity fluctuation of the particle image. For example, the image \textcolor{black}{of}  optically anisotropic, uniaxial particle observed between cross-polarizers appears darker or brighter, according to the angle formed between its optical axis and the axes of the polarizers. In this condition, rotational diffusion, by promoting the random reorientation of the particle, leads to an intermittent  "blinking" of the intensity associated to the particle image. If DDM analysis is now performed on an image sequence collected in these conditions, a $q$-independent decay is observed in $f(q,\tau)$, which is particularly evident in the low-$q$ regime. With proper fitting of $f(q,\tau)$ the characteristic correlation time of this decay can be measured and used to determine the associated rotational diffusion coefficient \cite{giavazzi2016simultaneous}. A similar effect, although realized \textit{via} a different optical mechanism, is exploited with dark-field microscopy to measure the characteristic reorientation time of micrometer-scale, shape-anisotropic particles \cite{cerbino2017dark}.

Using the same strategy for small anisotropic objects imaged in bright-field or in non-polarized fluorescence microscopy remains much more challenging: \textcolor{black}{an in-plane rotation of the object simply leads to a rotated image, while out-of-plane rotations can introduce subtle shape changes, but with no effect on the overall associated intensity.} In reciprocal space, this corresponds to the fact that the low-$q$ portion of the image power spectrum is not affected by rotations. In theory, rotational diffusion contributions become important in scattering at large wave-vectors $q$, where the object rotation can induce a fluctuation in the scattered intensity \cite{berne2000dynamic,pecora1968spectral}. However, the relaxation of the intermediate scattering function for such large $q$ is usually dominated by the translational term whose relaxation rate, in the case of a Brownian particle, rapidly grows as $D_Tq^2$, where $D_T$ is the translational diffusion coefficient.
To the best of our knowledge, the only study reporting a DDM-based measurement of the rotational dynamics of colloidal particles in bright-field is Ref. \cite{wittmeier2015rotational}, in which the particles translational motion was suppressed by partially tethering them to a solid surface. 
A general recipe for using bright field microscopy to simultaneously characterize the translational and rotational diffusive dynamics of Brownian particles remains thus unavailable.

In this work, we address this issue by combining the standard DDM analysis of bright-field movies with a simple image preprocessing step, whose function is to digitally reproduce the intensity change induced by particle rotation that is obtained by optical means in polarized or dark-field microscopy: for each image, we generate a local-orientation map whose intensity level encodes the local particle orientation at those image pixels where a particle is present; we then perform a standard DDM analysis on these local orientation maps and obtain the accurate characterization of both translational and rotational diffusion. A detailed description of the implementation of the method is reported in Sec. \ref{sec:SGDDM}, while a simple analytic justification is discussed in Appendix A.

\section{Materials and methods}
\subsection{Sample preparation and imaging \label{sec:sample}}
The sample is a highly diluted aqueous suspension of hematite-silica core-shell peanut-shaped particles, synthesized according to the protocol described in detailed Refs. \cite{sugimoto1993preparation,pal2018anomalous}. The length and diameter of the lobes of these particles are \textcolor{black}{$(1723 \pm 50)$ nm and $(740 \pm 50)$ nm}, respectively. Before measurement, the sample is loaded in a glass capillary of thickness $100$ $\mu m$, which is then sealed with vacuum grease. After about one hour at room temperature \textcolor{black}{($(T=22 \pm 1)$ $^o$C)}, the sample appears to be sedimented close to the bottom of the container \textcolor{black}{as the sedimentation length $l_g\sim 175$ $nm$ of these particles is smaller than their diameter \cite{kamal2020path}}. \textcolor{black}{The final number density is about $5\cdot 10^3$ $mm^{-2}$, as obtained from particle counting (see Fig. \ref{snapshot})}. 

The sample is observed with an inverted Nikon Ti-E bright-field microscope equipped with a Hamamatsu Orca Flash 4.0 camera (pixel size $6.5$ $\mu m$). The condenser diaphragm (numerical aperture NA=0.52) is kept completely open to achieve incoherent illumination of the sample. Two different objectives are used, a 40X, 0.60 NA objective and a 10X 0.25 NA objective. The size of the imaged region is $6.9 \cdot 10^3$ $\mu m^2$ and $2.8\cdot 10^4$ $\mu m^2$ for the higher and the lower magnification objective, respectively. Image sequences of $10^4$ frames each were recorded at 25 frame/s.

\subsection{Differential dynamic microscopy \label{sec:DDM}}
The translational dynamics of the particles is characterized by analyzing the sample movies with the standard DDM analysis scheme   \cite{cerbino2008differential,giavazzi2009scattering}.  
In short, we calculate the difference $\Delta I(\mathbf{x},t,\tau)=I(\mathbf{x},t+\tau)-I(\mathbf{x},t)$ between two images acquired at times $t$ and $t+\tau$. By averaging the spatial Fourier power spectrum of $\Delta I(\mathbf{x},t,\tau)$ obtained for the same $\tau$ but different reference times $t$ we obtain the so called \textit{image structure
function}
\begin{equation}
D(\mathbf{q},\tau)=\left\langle \left|\Delta\hat{I}(\mathbf{x},t,\tau)\right|^{2}\right\rangle _{t}\label{eq:struf}
\end{equation}
that captures the sample dynamics as a function of the two-dimensional
scattering wavevector $\mathbf{q}$ and of the lag time $\tau$. The symbol $\hat{\cdot}$ indicates the two dimensional digital Fourier transform, usually performed with a Fast Fourier Transform algorithm. 
In a last step, we take advantage of the circular symmetry of the problem to also perform an azimuthal average of $D(\mathbf{q},\tau)$, which leads to the one-dimensional function $D(q,\tau)$ of the radial wavevector
$q=\sqrt{q_{x}^{2}+q_{y}^{2}}$. 
The obtained structure function is connected to the intermediate scattering function (ISF) $f(q,\tau)$ \cite{berne2000dynamic} by the relation
\begin{equation}\label{STRUCTURE}
D(q,\tau)=A(q)\left[1- f(q,\tau)\right]+B(q)
\end{equation}
where the term $B(q)$ accounts for the camera noise and the amplitude $A(q)$ depends on the scattering properties of the sample and the transfer function of the imaging system \cite{giavazzi2014digital}.

\subsection{Squared-gradient differential dynamic microscopy \label{sec:SGDDM}}
\begin{figure}
\resizebox{1\columnwidth}{!}{%
  \includegraphics{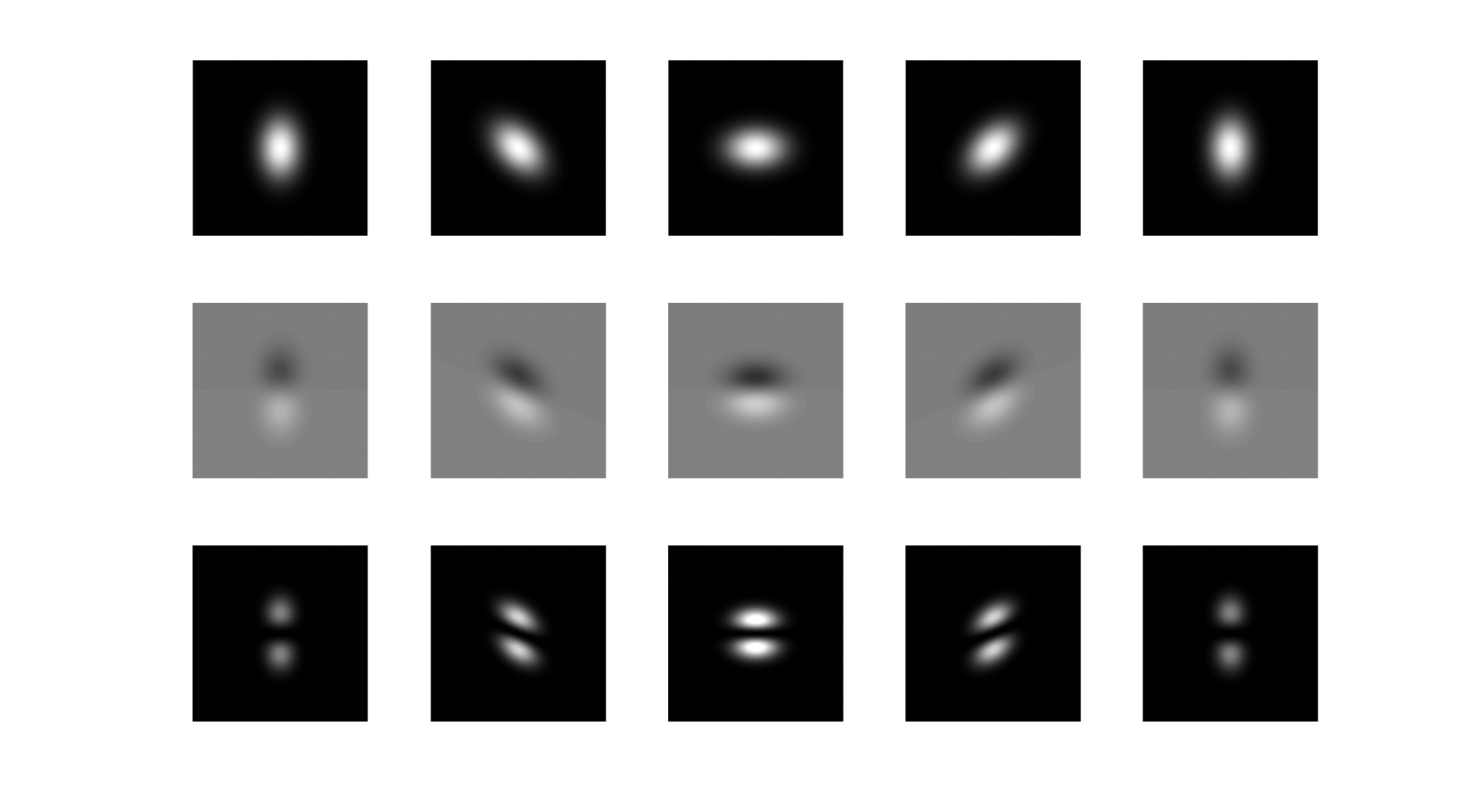}
}
\caption{\label{figura1}
"Squared-gradient" maps for a single anisotropic particle whose long axis lays on the image plane. First row: simulated images of the same particle with bivariate Gaussian intensity profile and different orientations. Second row: "gradient" maps. Each map correspond to the partial derivative along the vertical direction of the image above it. Third row: "squared-gradient" maps. Each map corresponds to the square of the map above it. In each row, maps are shown with the same grayscale.}
\end{figure}
In order to probe also the rotational dynamics of the particles we used a digital pre-processing step of the bright-field images to reveal information about the particles orientation. A simple implementation of this concept, which we call "squared-gradient" (SG), involves the calculation of the partial derivative of the image with respect to a given component $\mu$, which is then squared to give
\begin{equation}\label{SG}
c_\mu(\mathbf{x},t)=\left[\partial_\mu I(\mathbf{x},t)\right]^{2}.
\end{equation}
\textcolor{black}{While other implementations are also possible \cite{DIP}, in this work, the partial derivatives in Eq. \ref{SG} are calculated simply as the difference between the original image and its copy translated along one of the main axes by a single pixel}. As it can be appreciated in Fig. \ref{figura1}, where a single elongated particle is considered, the obtained map has the required property, \textcolor{black}{as the overall intensity changes according to the particle's orientation.
Indeed,} when the long axis of the particle is oriented along the $\mu$ axis (the vertical direction in Fig. \ref{figura1}), the contrast in the gradient map is lower compared to the case where the particle is oriented in the perpendicular direction. In the square-gradient map $c_\mu(\mathbf{x},t)$ the particle is replaced by two smaller lobes. The most evident effect of the particle rotation on $c_\mu(\mathbf{x},t)$ is a modulation of the intensity associated with the two lobes.
As shown in Appendix A, a similar intensity modulation in $c_\mu(\mathbf{x},t)$ is also produced by changes in the angle formed by the long axis of the particle with optical axis.


Once a sequence of SG-maps are obtained from the original images, the standard DDM analysis is performed to calculate the corresponding structure function
\begin{equation}
\label{energy}
D_{SG}(q,\tau)=\sum_{\mu=1,2}\langle|\hat{c}_\mu(\mathbf{q},t+\tau)-\hat{c}_\mu(\mathbf{q},t)|^2\rangle_{t,|\mathbf{q}|=q}.
\end{equation}
where we also average over the two orthogonal directions. \textcolor{black}{As it is shown explicitly in Appendix A for the special case $q=0$,} $D_{SG}(q,\tau)$ can be expressed in terms of the normalized ISF \textit{via} Eq. \ref{STRUCTURE} and we employ the general expression 
\begin{equation}\label{ISFS}
f(q,\tau)=\alpha(q)f_{RT}(q,\tau) + [1-\alpha(q)]f_{T}(q,\tau),
\end{equation}
where $f_T$ is the translational ISF, $f_{RT}$ is the roto-translational one and $\alpha(q)$ is the $q$-dependent relative amplitude of the roto-translational term. In Eq.\ref{ISFS}, we neglect a term describing the coupling between orientational and translational diffusion, which is expected to be of little significance \cite{han2006brownian}. For monodisperse Brownian particles, the ISFs take the well know expressions $f_T(q,\tau)=e^{-D_Tq^2\tau}$ and $f_{RT}(q,\tau)=e^{-6D_R\tau-D_Tq^2\tau}$, where $D_T$ and $D_R$ are the average translational and rotational diffusion coefficients, respectively \cite{berne2000dynamic}. \textcolor{black}{It is worth stressing that the taking the square in Eq. \ref{SG} represents an essential step of the procedure, as it enables "translating" the orientation-dependent spatial modulation introduced by the gradient operation into a global intensity variation, which affects in a $q$-independent fashion all the Fourier components of the SG map.}
\subsection{Single particle tracking \label{sec:SPT}}
To validate SG-DDM, we analyze the same images also with single particle tracking (SPT), by using the MATLAB code developed by the Kilfoil group at the University of Massachusetts \cite{PhysRevLett.102.188303} and available at https://github.com/dsseara/microrheology. We reconstruct individual trajectories of the center of mass of the particles and we calculate their mean-squared displacement $\langle \Delta r^2(\tau) \rangle$ as a function of the delay time $\tau$.

To track over time the orientation of the particles, we use a custom MATLAB code \textcolor{black}{implementing a procedure similar to the one described in Refs. \cite{PhysRevE.80.011403,zheng2010self}, which is described in detail in Appendix B}.
We applied the described procedure, which provides an independent estimate of the rotational diffusion coefficient, only to the image sequence recorded with the higher magnification (40X), as we were not able to successfully perform the same orientational tracking-based analysis on the image sequences acquired with a 10X objective. 

\section{Results and discussion}

We show in Fig. \ref{ISFs} representative ISFs obtained from both standard DDM (panel a) and SG-DDM (panel b) analysis of the same sequence of bright-field images (magnification 40X) of the sample. The ISFs obtained from standard DDM analysis exhibit a single exponential decay $e^{-\Gamma(q)\tau}$, whose relaxation rate $\Gamma(q)$ displays a clean $\sim q^2$ scaling (Fig. \ref{fig1}, gray triangles).
\begin{figure}
\resizebox{1\columnwidth}{!}{%
  \includegraphics{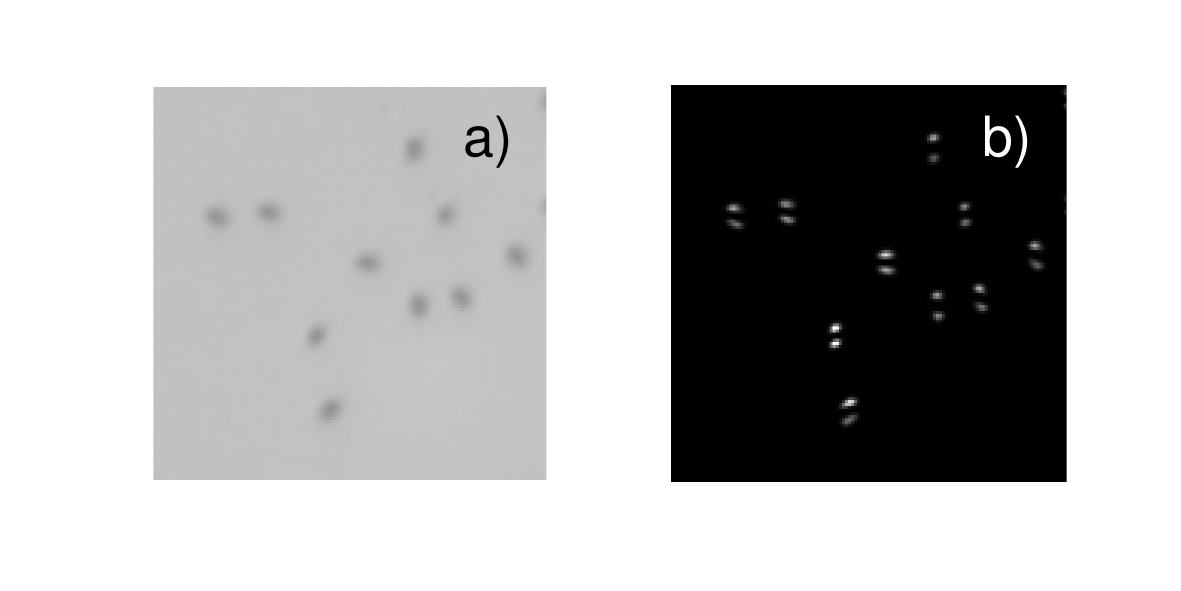}
}

\caption{\label{snapshot} a) Representative bright-field image of anisotropic peanut-shaped particles diffusing close to the bottom of the glass capillary (see text for details) \textcolor{black}{taken with a 40X, 0.60 NA objective}. The size of the image corresponds to 39 $\mu$m in real space.
b) Square-gradient map obtained from the image shown in panel (a). Each particle appears to be replaced by two bright spots whose intensity depends on the particle orientation.}
\end{figure}

 A fit of $\Gamma(q)$ with a quadratic model $\Gamma(q)=D_{T}q^2$ provides an estimate for the translational diffusion coefficient $D_T=0.24 \pm 0.02$ $\mu m^2/s$ of the particles, which is in excellent agreement with the value $D_{T,PT}=0.23 \pm 0.01$ $\mu m^2/s$ obtained with SPT (Fig. \ref{PT}a) \textit{via} a linear fit $\langle \Delta r^2(\tau) \rangle = 4 D_{T} \tau$ of the MSD of the particles.
 

\begin{figure}
\centering
\resizebox{0.75\columnwidth}{!}{%
  \includegraphics{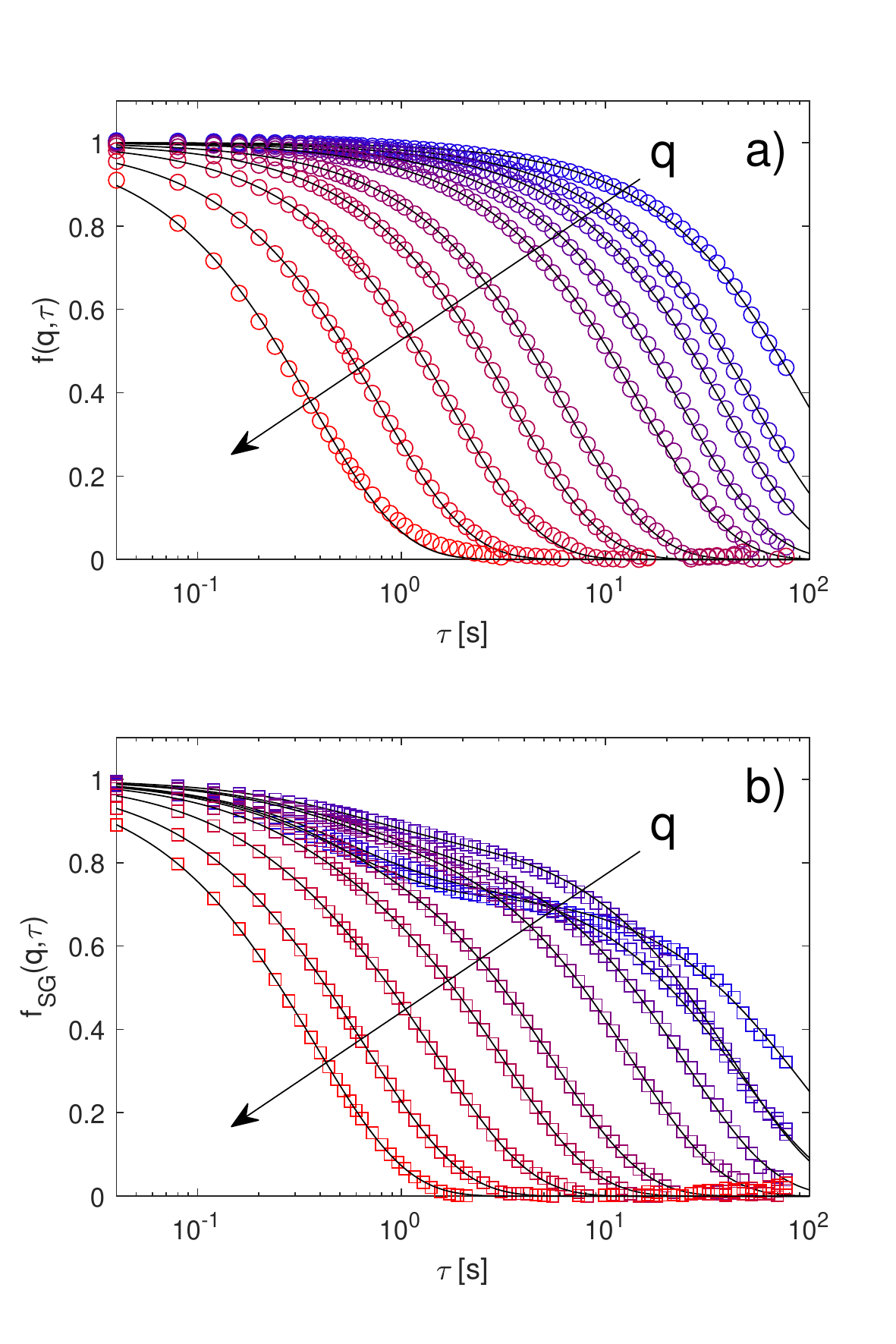}}
\caption{\label{ISFs}
a) Representative ISFs obtained from standard DDM analysis of an image sequence of Brownian anisotropic particles for different $q$-values, logarithmically spaced in the interval $[0.15, 3.4]$ $\mu m^{-1}$. Continous black curves are best fitting curves to the data with a simple exponential model. b) Representative ISFs obtained from SG-DDM analysis of the same image sequence and for the same $q$-values considered in panel (a). Continous black curves are best fitting curves to the data with a double exponential model.}
\end{figure}

Inspection of the ISFs obtained from SG-DDM (Fig. \ref{ISFs}b) reveals instead the existence of a double decay, which is particularly evident for the lowest values of $q$. Fitting a double exponential model to the data allows to reliably estimate, at least for $q\lesssim2.5$ $\mu m^{-1}$, two well distinct decay rates: a fast decay with an almost $q$-indpendent rate $\Gamma_1$, and a slow diffusive decay with rate $\Gamma_2\sim q^2$ that is fully compatible with the relaxation rate obtained from standard DDM (Fig. \ref{fig1}). For $q>2.5$ $\mu m^{-1}$, the two relaxation rates are too close to be reliably separated and the fitting procedure provides a single relaxation rate.

According the discussion in Sec.\ref{sec:SGDDM}, we attribute the fast decay to the roto-traslational diffusion of the particles, $\Gamma_1(q)=6D_R+D_T q^2$, and we obtain the estimate $D_R=0.30 \pm 0.05$ $s^{-1} $. This result agrees with those obtained with SPT according to the procedures described in Section \ref{sec:SPT}, in particular with the estimate $D_{R,PT_1}=0.28 \pm 0.05$ $s^{-1}$  obtained from a linear fit of the angular MSD $\langle \Delta \phi^2(\tau) \rangle = 2 D_{R,PT_1} \tau$ (Fig. \ref{PT}b), and with the estimate $D_{R,PT_2}=0.29 \pm 0.05$ $s^{-1}$  obtained from an exponential fit of the effective aspect ratio autocorrelation function $C_{\epsilon}(\tau)=e^{-6D_{R,PT_2}\tau}$ (Fig. \ref{PT}c).

We perform the same analyses on an image sequence of the same sample, acquired with a 10X objective (NA=0.25). In this imaging condition, the particle size is very close to the diffraction limit ($d\sim \overline{\lambda}/(2NA)\simeq 1.2$ $\mu m$, assuming $\overline{\lambda}\simeq 0.6$ $\mu m$) and the effect of the convolution with the point-spread-function is to markedly reduce the apparent anisotropy in the particle image, that is now barely appreciable (inset of Fig. \ref{fig1}). As a consequence, the uncertainty in the frame-by-frame determination of single particle orientation is so large that the 
angular tracking procedure described in Section \ref{sec:SPT} no longer provides reliable results. On the contrary, the results of both DDM and SG-DDM are in excellent agreement with the one obtained with the 40X objective (NA=0.60), as shown in Fig. \ref{fig1}.

\textcolor{black}{
As a further validity check for the proposed approach, we compare the obtained diffusion coefficients with the ones predicted by the theory for a rigid rod, which represents a fair model for our peanut-shape particles. If we consider a cylinder of length $1723$ nm and diameter $740$ nm, in a Newtonian fluid of viscosity $\eta=1.0$ mPa at temperature $T=22$ $^o$C, by using the theoretical expressions reported in Ref. \cite{PhysRevE.50.1232} (Eqs. 1-4) we obtain the following values  
. $D_{R,th}=0.44$ $s^{-1}$, $D_{T,th}=0.25\cdot$ $\mu m^2/s$ for the rotational and the translational diffusion coefficient, respectively. We note that, while $D_{T,th}$ is in very good agreement with the corresponding values obtained with both SG-DDM and SPT, the calculated diffusion coefficient $D_{R,th}$ is off by about 30\% with respect the experimental ones. This moderate discrepancy could be explained as a systematic effect of the adopted cylindrical approximation. However, it can be also a genuine effect due the hydrodynamic interactions between particles and the bottom plate of the container, which can slow down the particles' diffusivity compared to the bulk \cite{Yang}.
}  

\begin{figure}
\centering
\resizebox{0.75\columnwidth}{!}{%
  \includegraphics{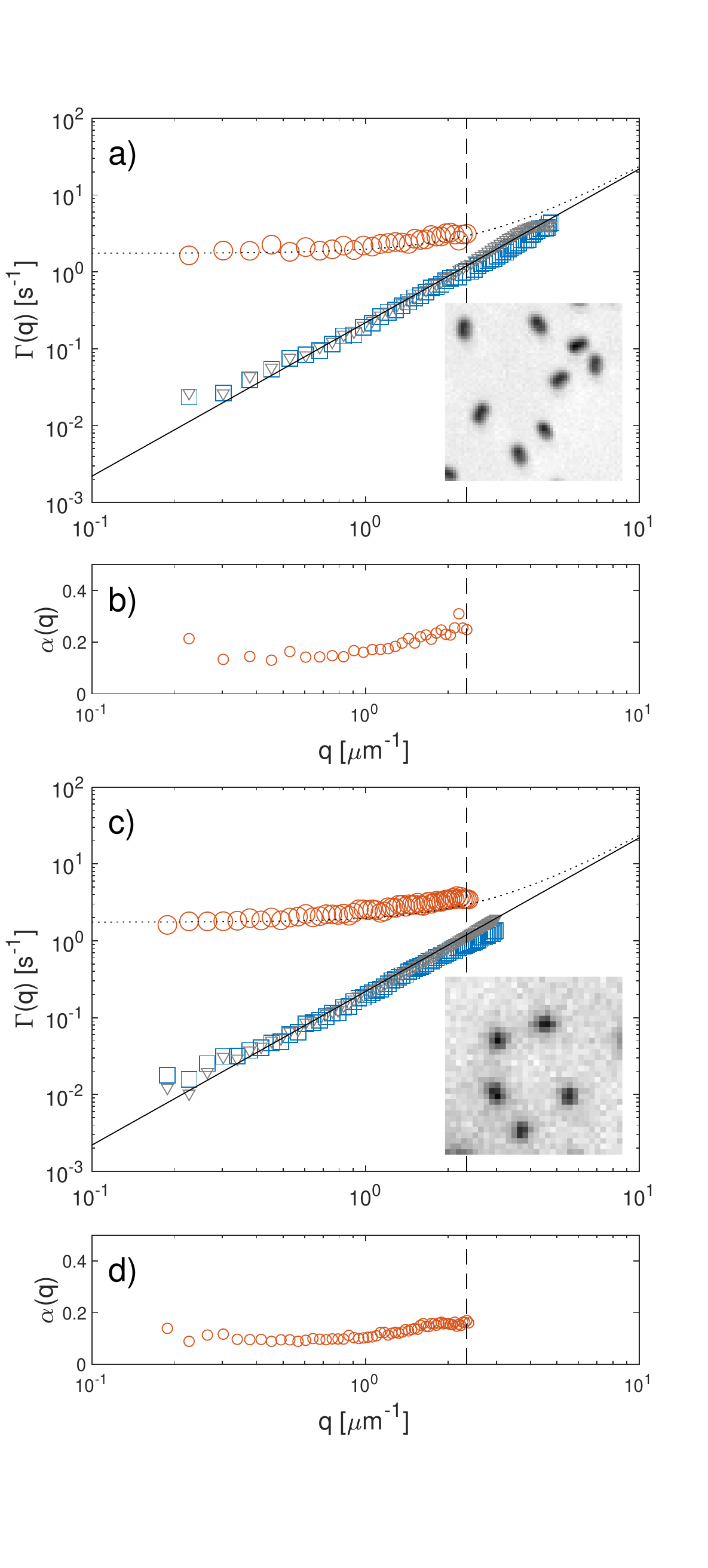}}
\caption{\label{fig1} Roto-translational dynamics of anisotropic Brownian particles obtained from DDM and SG-DDM. Panels (a-b) and (c-d) show results obtained from the analysis of image sequences of the same sample, acquired with a 40X, 0.60 NA objective and a 10X, 0.25 NA objective, respectively. 
a) Gray triangles: relaxation rate $\Gamma(q)$ obtained from the single exponential fit of the structure functions obtained from standard DDM analysis. Orange circles and blue squares: relaxation rates $\Gamma_1(q)$ and $\Gamma_2(q)$ obtained from a double exponential fit of the structure functions obtained from SG-DDM analysis on the same image sequence. The continous line $\Gamma=D_Tq^2$ is a best fitting curve to the blue squares. The analogous curve obtained from the gray triangle is indistinguishable.
The dashed curve is a best fitting curve $\Gamma_1=6D_R+D_Tq^2$ to the orange circles. The vertical dashed line marks the $q$-value above which the two relaxation rates $\Gamma_1$ and $\Gamma_2$ cannot be resolved any more. The small inset shows a representative ROI of size 10.4 $\mu m$. \textcolor{black}{b) Orange circles: relative amplitude $\alpha(q)$ of the roto-translational contribution to the ISF (see Eq.\ref{ISFS}), obtained from a double exponential fit of the SG-DDM structure functions. The description of panels c) and d) is identical to the one of panels a) and b), respectively}}
\end{figure}

\begin{figure*}
\resizebox{2\columnwidth}{!}{%
  \includegraphics{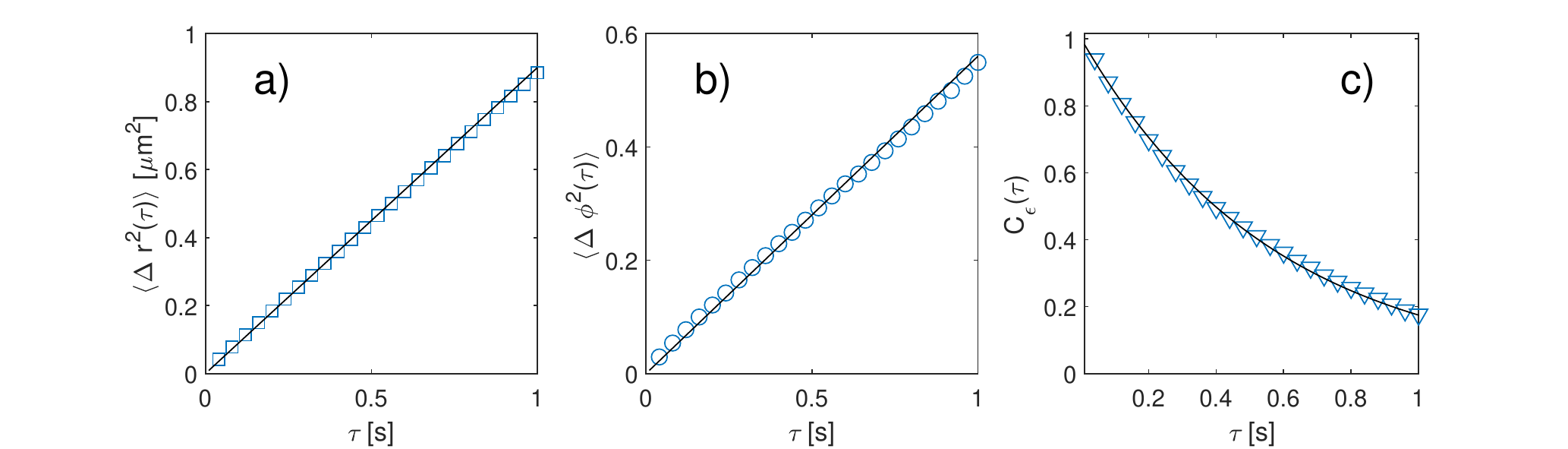}}
\caption{\label{PT} Roto-translational dynamics of anisotropic Brownian particles obtained from single particle tracking. a) Symbols: MSD of the center of mass of the particles. The continous line a best fitting curve to the data with a linear model $\langle \Delta r^2(\tau) \rangle = 4 D_{T} \tau$, with $D_{T}=(0.23\pm 0.01)$ $\mu m^2s^{-1}$ b) Symbols: polar angular MSD of the particles. The continuous line a best fitting curve to the data with a linear model $\langle \Delta \phi^2(\tau) \rangle =2 D_{R} \tau $, with $D_{R}=(0.28 \pm 0.05)$ $s^{-1}$ . c) Symbols: normalized temporal correlation function of the apparent aspect ratio $\epsilon(t)=b^2(t)/a^2(t)$ of each particle, as obtained from a bivariate Gaussian fit of the associated intensity distribution (see main text for details). The continuous line a best fitting curve to the data with an exponential decay $ C_\epsilon(\tau)=\exp(-\gamma \tau) $, with $\gamma=(1.75 \pm 0.2)$ $s^{-1}$ }
\end{figure*}

\section{Conclusions}
In this work, we introduced SG-DDM, a simple procedure enabling the tracking-free determination of the rotational and translational dynamics of anisotropic particles in microscopy images. SG-DDM was here demonstrated for bright-field microscopy experiments, but we trust that it could be applied without modifications to a broad range of imaging modes, including wide-field or confocal fluorescence microscopy.

Compared to SPT-based approaches, we highlight a number of significant advantages. First of all, at a variance with most SPT methods, the proposed approach does not require user-dependent parameters and does not rely on a specific model for the particle shape. Moreover, by exploiting the intrinsic ensemble-averaging capability of the method, we proved that SG-DDM is able to provide a quantitative information on the rotational dynamics even when the size of the particles is very close to the diffraction limit and a low signal-to-noise ratio does not allow a reliable frame-by-frame determination of single-particle orientation in real space.
\textcolor{black}{We thus expected the method to be compatible with a wide range of shape anisotropic particles, such as for example metallic nanowires \cite{farain2018universal}, as well as with spherical,  but optically anisotropic, objects, like Janus particles \cite{wittmeier2015rotational}}.
Computationally, SG-DDM is as effective as DDM, whose use becomes advantageous in particular when a large number of particles are considered. In fact, the computation time does not depend on the number of imaged particles, so that thousands of particles can be analyzed in parallel in a single movie.

Being based on a non-linear pre-processing step, a condition under which the method is expected to provide the correct translational dynamics of the sample is the absence of a significant overlap between images of the different particles \cite{giavazzi2009scattering,cerbino2017dark}. While this condition is almost automatically fulfilled for rigid objects in two dimensions, it could represent a more serious limitation on the particle concentration in the three dimensional case. 

\textcolor{black}{In addition, if the particles deviate significantly from a simple uniaxial symmetry, such as the one investigated here, higher other contributions (corresponding to high order terms in the spherical harmonics expansion of the particles’ shape) become relevant in the ISF. In those cases, a more refined system-dependent model for the ISF, including multiple exponential decays, should be used to properly interpret the SC-DDM signal.}

\textcolor{black}{Finally, also }the presence of a strongly inhomogeneous optical background can affect the quality of the results, as the non-linear processing can introduce a coupling between moving objects of interest and static features. A mitigation of this problem could be provided by a background subtraction before calculating the SG-map. For example, if the observation time window is long enough to enable a complete decorrelation in the particles positions, a good estimate of the background intensity could be obtained as the time average of the whole sequence.

\textcolor{black}{In all the above mentioned cases, }a quite compelling internal quality test is provided by the comparison between the results obtained from standard DDM and SG-DDM on the same sequence, as the exact same translational dynamics is expected to be measured in DDM and SG-DDM. Any lack of correspondence between the translational ISFs obtained with the two methods is a proxy of a potential artifact introduced by the SG mapping, offering a simple and sensitive check of the validity of the mapping procedure.

Finally, beyond offering a novel way to determine the rotational dynamics of small anisotropic entities, SG-DDM represents also the first example of non-linear pre-processing of microscope images prior to DDM analysis, which we believe will expand the range of applications of DDM and other DFM techniques beyond the current frontier.

\section*{Appendix A. Analytical calculation of $D_{SG}(q=0,\tau)$ \label{coeffs} }

In this Appendix we provide analytical justification of SG-DDM as a probe of the rotational diffusion of anisotropic particles. In particular, we show that, in the $q\rightarrow 0$ limit, the structure function $D_{SG}(q,\tau)$ is dominated by a $q$-independent relaxation mirroring the rotational dynamics. 

We assume a simple model where each "particle" described by a 3D Gaussian profile with variances $(a^2,a^2,b^2)$ along its three main axes ($a<b$). Let $(x_0,y_0,z_0)$ be the coordinate of its center of mass, while the orientation of the long axis is determined by the polar angle $\theta$ (\textit{i.e.} the angle formed with the $z$-axis, corresponding to the optical axis) and the azimuthal angle $\phi$.

In what follows, we consider for simplicity the case of a single particle, the generalization to a collection of identical, non overlapping, particles being straightforward.
We assume that the intensity distribution $I(x,y)$ on the image plane is given by the projection on the $x-y$ plane of the object density distribution (good for example in the case of incoherent bright-field or wide-field fluorescence microscopy of a particle close to the object plane \cite{giavazzi2009scattering}). Under this hypothesis, $I(x,y)$ is a 2D Gaussian distribution $I(x,y)=1/(\sigma_x\sigma_y)\exp \left(-{x'^2}/{2\sigma_x^2} -{y'^2}/{2\sigma_y^2} \right)$, where $x'=(x-x_0)\cos\phi+(y-y_0)\sin\phi$, $y'=(y-y_0)\cos\phi-(x-x_0)\sin\phi$, $\sigma_x=a$ and $\sigma_y^2=a^2\cos^2\theta+b^2\sin^2\theta$.

The total intensity $i_0=\int d^2\mathbf{x} c_\mu(\mathbf{x},t)=\hat{c}_\mu(\mathbf{q}=0,t)$ associated with the gradient map $c_\mu(\mathbf{x},t)$ can be explicitly calculated. We find  $i_0(\theta,\phi)=c\left(\frac{\sigma_x^2+\sigma_y^2}{\sigma_x^3\sigma_x^3}\right)\left[1+\left(\frac{\sigma_y^2-\sigma_x^2}{\sigma_x^2+\sigma_y^2}\right)\cos2\phi\right]$, where $c$ is a dimensionless constant.

We initially consider the case of a moderately anisotropic particle (\textit{i.e.} one whose eccentricity $e\equiv \sqrt{1-a^2/b^2}$ is small), for which we obtain the following expression, valid to the first order in $e$:
\begin{equation}\label{i0}
i_0(\theta,\phi)=c'\left[1+e^2\sin^2\theta\left(1+(1/2)\cos2\phi \right)\right],
\end{equation}
where  $c'=(2c/a^4)$.

It is worth expressing $i_0$ in terms of the spherical harmonics $Y_l^m(\theta,\phi)$
\begin{equation}\label{i000}
i_0(\theta,\phi)=a_{00}Y_0^0+a_{20}Y_2^0+a_{22}[Y_2^{-2}+Y_2^{2}],
\end{equation}
where 
$a_{00}=2\sqrt{2}(1+2e^2/3)$, $a_{20}=(4/3)e^2\sqrt{\pi/5}$ and
$a_{22}=e^2\sqrt{2\pi/15}$.

Combining Eq. \ref{i000} with the identity \cite{berne2000dynamic}
\begin{equation}
\langle Y_l^{m}Y_{l'}^{m'} \rangle=\frac{1}{4\pi}e^{-l(l+1)D_R\tau}\delta_{l,l'}\delta_{m,m'},
\end{equation}
we obtain the following expression for $D_{SG}(q=0,\tau)=2\left(\langle i^2_0(t)\rangle - \langle i_0(t)i_0(t+\tau) \rangle  \right)$
\begin{equation}
\label{eq:Dq0}
D(q=0,\tau)=A_0(1-e^{-6D_r\tau}).
\end{equation}
This simple expression is valid under the hypothesis of small shape anisotropy. Within the same approximation, the effective aspect ratio $\epsilon=\sigma_y/\sigma_x$ is given by  $\epsilon= 1 + \frac{1}{2}e^2\sin^2\theta$ and the corresponding time autocorrelation function $C_{\epsilon}(\tau)=\left[\langle \epsilon(t+\tau)\epsilon(t) \rangle -\langle \epsilon(t) \rangle^2 \right]/\left[ \langle\epsilon(t)^2\rangle -\langle\epsilon(t)\rangle^2 \right]$ can be easily calculated as

\begin{equation}
\label{eq:Ceps}
C_{\epsilon}(\tau)=e^{-6D_r\tau}.
\end{equation}

If the eccentricity is large, higher order spherical harmonics are expected to appear in Eq.\ref{i000}, leading to multiple exponential decay of the correlation functions \cite{berne2000dynamic}.

\section*{Appendix B. Rotational tracking algorithm}

\begin{figure}
\centering
\resizebox{.75\columnwidth}{!}{%
  \includegraphics{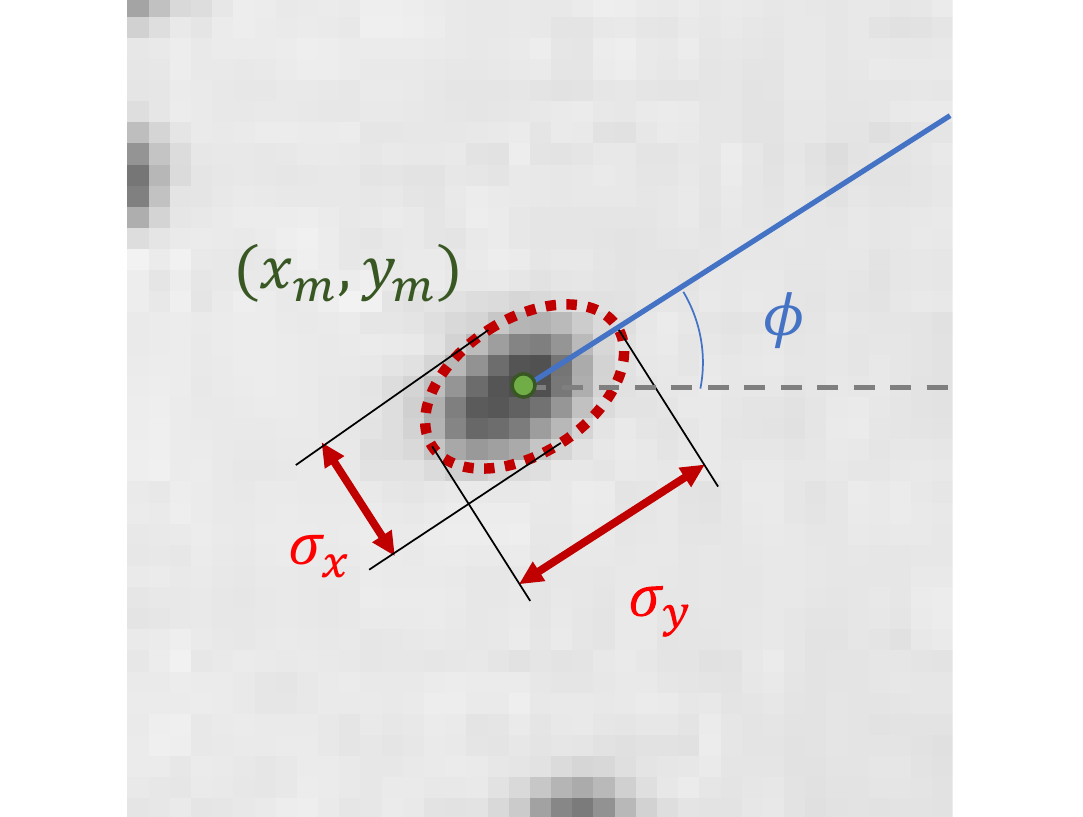}
}
\caption{\label{figurarottrack} \textcolor{black}{Schematic representation of the rotational tracking algorithm. The coordinates $(x_m,y_m)$ of the centroid of each particle are determined $via$ a standard particle tracking algorithm \cite{PhysRevLett.102.188303}. The shape and orientation of the particle's image are determined \textit{via} a bivariate Gaussian fit to the local intensity profile. This enables estimating the length $\sigma_x$ and $\sigma_y$ of the short and of the long axis, respectively, and the azimuthal angle $\phi$ formed by the particle's long axis with the horizontal direction. The effective aspect ratio $\epsilon=\sigma_y/\sigma_x$ depends of the polar angle $\theta$ between the particle's axis and the vertical direction, as described in the text.}
}
\end{figure}
\begin{figure}
\resizebox{1\columnwidth}{!}{%
  \includegraphics{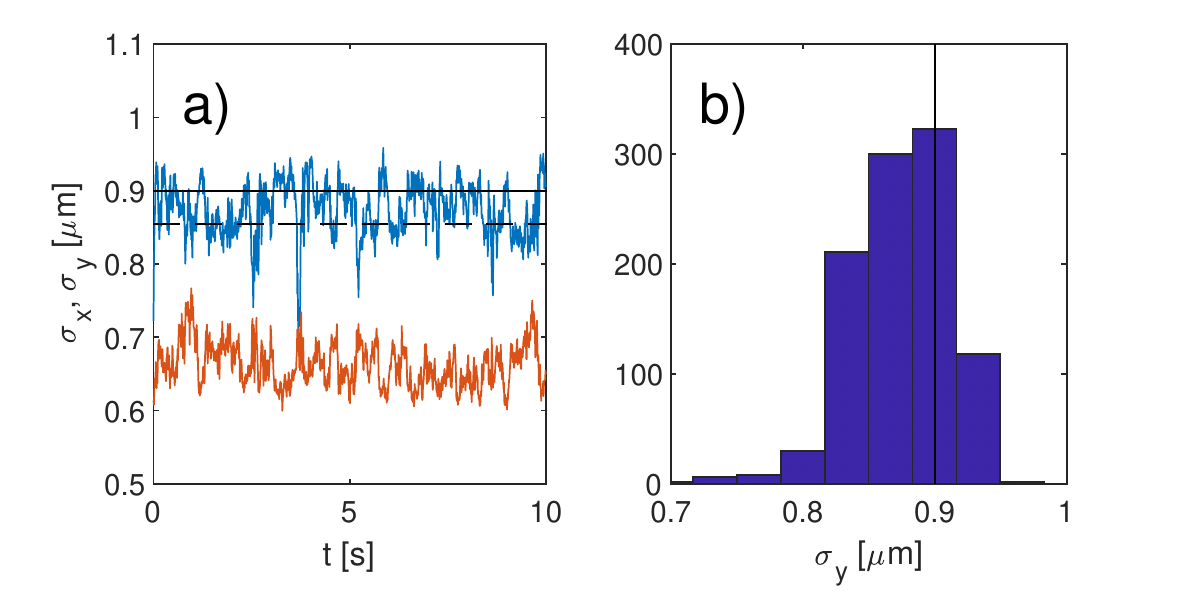}
}
\caption{\label{figurarrottrack} a) Blue (orange) curve: time-dependent length of the major (minor) axis $\sigma_y(t)$ ($\sigma_x(t)$) of a single particle, as obtained from a bivariate Gaussian fit of its intensity distribution. The black continuous line corresponds to the mode $\overline{\sigma_y}$ of the frequency distribution of $\sigma_y(t)$}, while the black dashed line corresponds to threshold value $0.95 \overline{\sigma_y}$ (see main text for details). b) Frequency distribution of  $\sigma_y(t)$. The vertical line corresponds to the mode $\overline{\sigma_y}=(0.90 \pm 0.02)$ $\mu m$ of the distribution.
\end{figure}

\textcolor{black}{
 In this Appendix, we present a detailed description of the algorithm used to track the rotational motion of the particles in real space. As described in Par. \ref{sec:SPT}, we first apply a standard particle tracking algorithm \cite{PhysRevLett.102.188303} to determine, frame by frame, the position of the centroid of each particle.} For each tracked particle $i$ and for each time $t$, a square ROI of size 16x16 pixel \textcolor{black}{(corresponding to area of about 27 $\mu m^2$ when the 40X objective is used)} is obtained from the original image, centered on the particle's centroid; a bivariate Gaussian distribution with arbitrary orientation of the main axes
$ f(x,y)=A\exp{-[x'^2/2\sigma_x^2-y'^2/2\sigma_y^2]}$, with 
$x'=[(x-x_m)\cos\phi+(y-y_m)\sin\phi]$ and $y'=[(x-x_m)\sin\phi-(y-y_m)\cos\phi]$, is fitted to the ROI intensity distribution, which provides an estimate for the variances $\sigma_x^2(t)$ and $\sigma_y^2(t)$ along the main axes, and for the azimuthal angle $\phi_i(t)$ identifying the direction of projection of the long axis on the $x,y$ plane. The values obtained for $\sigma_x(t)$ and $\sigma_y(t)$ are quite consistent along each trajectory and display time-correlated fluctuation around their mean values, which we attribute to random variations in the polar angle $\theta$ formed by the long axis of the particle with the vertical direction (see Fig. \ref{figurarrottrack}).

We then calculate the autocorrelation function $C_{\epsilon}(\tau)=\left[\langle \epsilon(t+\tau)\epsilon(t) \rangle -\langle \epsilon(t) \rangle^2 \right]/\left[ \langle\epsilon(t)^2\rangle -\langle\epsilon(t)\rangle^2 \right]$ of the effective aspect ratio $\epsilon(t)=\sigma_y(t)/\sigma_x(t)$. For an elongated particle performing rotational Brownian motion, $C_{\epsilon}(\tau)$ is expected to display a simple exponential relaxation $C_{\epsilon}(\tau)=e^{-\gamma\tau}$, with $\gamma=6D_R$, as shown in Appendix A. 
This provides a robust means to measure the  rotational dynamics of the particles, namely by fitting an exponential function to the obtained $C_{\epsilon}(\tau)$.

In order to extract a similar information from the azimuthal degree of freedom, which is encoded in $\phi$, the simultaneous knowledge of the polar angle $\theta$ is also needed. This is due to the fact that the same change in the azimuthal angle $\delta\phi$ can correspond to actual displacements of very different amplitudes according to the value of $\theta$ \cite{anthony2015tracking}. To account for this, a different variable $\psi$ is often considered, which is defined by the relation $d\psi=d\phi\sin\theta$ \cite{colin2014rotational}. For a particle performing rotational diffusion, the mean squared value of  $\psi$ follows the simple relation $\langle\Delta\psi^2(\tau)\rangle=2D_R\tau$.

In our experiments, although we could in principle calculate $\theta(t)$ from $\sigma_x(t)$, $\sigma_y(t)$, we found that this direct inversion was too noisy to provide reliable results. 
As a consequence, we adopt a simplified yet more robust approach. For each particle, we consider the histogram of the values assumed by the long axis length $\sigma_y(t)$ and we identify the value $\overline{\sigma_y}$ corresponding to the right peak of the distribution (see Fig.\ref{figurarottrack}). 
According to the model described in the Appendix, $\overline{\sigma_y}$ corresponds to the length of the long axis when it lies on the $x-y$ plane, \textit{i.e.} when $\theta=\pi/2$.
For each particle, we identify the portions of the trajectory such that
\begin{equation}\label{dulabadula}
(\overline{\sigma_y}-\sigma_y(t))/\overline{\sigma_y}<0.05.
\end{equation}
This corresponds to $ |\theta_i(t)-\pi/2| \lesssim \pi/8$. Within each of these portions, $\Delta\phi\simeq \Delta \psi$ and thus $\langle\Delta\phi^2(\tau)\rangle\simeq2D_R\tau$, where the average is performed only over those portions for which Eq. \ref{dulabadula} is satisfied.

\section*{Acknowledgments}
We acknowledge funding from from the Associazione Italiana per la Ricerca sul Cancro (AIRC) - Project MFAG \# 22083, and from the Italian Ministry of University and Scientific Research (MIUR) - Project RBFR125H0M.

\section*{Author contribution statement}
FG and RC designed research. FG performed experiments and analyzed the data. AP synthesized the particles and prepared the samples. All authors discussed the experimental results. FG and RC wrote the paper.

\bibliography{biblio_x}
\bibliographystyle{epj}
 
%

%
%

\end{document}